\documentclass[fleqn,usenatbib,useAMS,usedcolumn]{mnras}

\usepackage{newtxtext}
\usepackage[varg]{newtxmath}
\usepackage[T1]{fontenc}
\usepackage{bm}		% Bold maths symbols, including upright Greek
\usepackage{graphicx}	% Including figure files
\title[Astrometric detectability of unseen companions]{Astrometric detectability of systems with unseen companions: effects of the Earth orbital motion}

\author[A. G. Butkevich]{
Alexey G. Butkevich\thanks{E-mail: ag.butkevich@gmail.com}
\\
Pulkovo Observatory, Pulkovskoye chaussee 65, Saint Petersburg 196140, Russia
}

\date{Accepted. Received; in original form}

\pubyear{2018}

\begin{document}
\label{firstpage}
\pagerange{\pageref{firstpage}--\pageref{lastpage}}
\maketitle

\begin{abstract}
Astrometric detection of an unseen companion is based on analysis of apparent motion of its host star around the system's barycentre. Systems with orbital period close to one year may escape detection if orbital motion of their host stars are observationally indistinguishable from parallax effect. Additionally, the astrometric solution may produce a biased parallax estimation for such systems. We examine effects of orbital motion of the Earth on astrometric detectability in terms of correlation between the Earth's orbital position and position of the star relative to its system barycentre. The $\chi^2$ statistic for parallax estimation is calculated analytically, leading to expressions that relate the decrease in detectability and accompanying parallax bias to the position correlation function. The impact of the Earth's motion critically depends on the exoplanet's orbital period, diminishing rapidly as the period deviates from one year. Selection effects against one-year period systems is therefore expected. Statistical estimation shows that the corresponding loss of sensitivity results in a typical 10 per cent increase in detection threshold. Consideration of eccentric orbits shows that the Earth's motion has no effect on detectability for $e\ga0.5$. Dependence of detectability on other parameters, such as orbital phases and inclination of the orbit plane to the ecliptic, are smooth and monotonic because they are described by simple trigonometric functions.
\end{abstract}


%%%%%%%%%%%%%%%%%%%%%%%%%%%%%%%%%%%%%%%%%%%%%%%%%%

%%%%%%%%%%%%%%%%% BODY OF PAPER %%%%%%%%%%%%%%%%%%

\section{Introduction}

Astrometric discovery of unseen companions relies on detection of the effects they exert on their host stars. \citet{sozzetti2013} gives a concise and informative introduction to this subject with an emphasis on the potential of microarcsecond astrometry and \cite{Perryman2011} provides a more detailed exposition of the astrometric technique in the context of exoplanetary studies.

The prospects for astrometric detection of exoplanets has greatly improved with the launch of the European Space Agency's space astrometry mission \emph{Gaia} aimed at accuracies at the 10 microarcsecond level \citep{Gaia-Collaboration2016}. A number of studies have been carried out to assess the \emph{Gaia} potential for discovering exoplanetary systems. \cite{Casertano+2008} demonstrated in their elaborate test program that \emph{Gaia} could reliably detect planets with orbital periods up to 5\,yr and astrometric signal-to-noise ratio ($\rmn{S}/\rmn{N})\gid3$. \cite{Perryman+2014} concluded that \emph{Gaia} should detect $\sim$20000 long-period giant planets ($M_\rmn{p}>1-15M_J$) out to distances of $\sim$500~pc. Considering population of nearby M dwarfs, \cite{Sozzetti+2014} showed that roughly 100 giant planet are detectable around these stars within 30\,pc. \cite{Sahlmann+2015} argued that discovery of several hundreds giant circumbinary planets is to be expected within 200\,pc. Exploring different detection criteria, \cite{2017arXiv1704.02493} found that detection level of $\rmn{S}/\rmn{N}\ga1-1.5$  is achievable for the nominal mission length of 5 years.

The aim of the present work is to consider the effects of the Earth orbital motion on astrometric detectability of an unseen companion, which are significant for systems with orbital period close to one year. It is clear from simple geometrical considerations that in such systems orbital motion of the host star, under certain conditions, may be observationally close to the parallactic effect or even indistinguishable from it. It means that the orbital motion may be partially or fully absorbed by the parallax parameters. This has two consequences. Firstly, the astrometric solution provides a biased parallax estimation. Secondly, the apparent motion of the host star is to the first order described by a single-star model, whereas astrometric detection of the unseen companion is based on a deviation from this model, hence the detectability of these companions decreases for such systems. These effects, parallax bias and loss in detectability, first demonstrated by \cite{Holl2011} and mentioned by \cite{Perryman+2014}, have not been systematically studied so far.

We discussed the coupling between orbital motion and parallax effect on the basis of goodness-of-fit $\chi^2$ statistic for parallax estimation. This approach leads to considerable simplification of the problem. Enabling us to easily calculate the parallax bias, it at the same time provides assessment of deviation from the single-star model. Moreover, it allows us to describe all the effects compactly in terms of one parameter $\rho$ (Eq.~(\ref{eq:corrfunc-def})), correlation between orbital positions of the Earth and host star. It is worth mentioning that this approach throws an interesting sidelight on the concept of astrometric signature, directly linking it to minimum value of $\chi^2$. If no coupling between orbital motion and parallax exists, such statistical definition results in the conventional expression for astrometric signature.

The present work mainly deals with astrometric detectability of substellar objects, i.e. planets and brown dwarfs. For brevity, we therefore refer to all such objects as planets hereafter. Similarly, by a planetary system we mean any system, in which a substellar companion orbits the host star. To avoid ambiguity, we shall explicitly indicate when a companion of stellar mass is discussed.

The outline of the paper is as follows. Section~\ref{s:earth} gives derivation of the parallax bias and loss in sensitivity in terms of the position correlation function. Section~\ref{s:application} contains an analytical treatment of the correlation function. Effects of various factors on the sensitivity are discussed in Sect.~\ref{s:discussion}. The conclusions are presented in Sect.~\ref{s:conclusions}.

\section{Effect of the Earth motion}\label{s:earth}

The standard astrometric model is based on the assumption that stars move uniformly relative to the solar system barycentre. In the framework of this model, the direction to a star as seen by an observer at barycentric position $\bmath{b}$ at time $t$ is given by the unit vector \citep[][Eq.~(4)]{lindegrenAGIS2012}
\begin{equation}\label{eq:uStar}
 \bmath{u}\left(t\right)=
 \left\langle\left\langle\bmath{r}+\left(\bmath{r}\mu_r+\bmath{\mu}\right)\left(t-t_0\right)-
 \varpi\bmath{b}\left(t\right)/A\right\rangle\right\rangle,
\end{equation}
where the unit vector $\bmath{r}$ specifies the barycentric direction of the star at initial time $t_0$, $\bmath{\mu}\left(\equiv\dot{\bmath{r}}\right)$ is the proper motion vector, and the radial proper motion $\mu_r$ is calculated from the trigonometric parallax  $\varpi$ and the radial velocity $v_r$ as $\mu_r=v_r\varpi/A$, with $A$ being the astronomical unit. The double angular brackets denote vector normalization, $\left\langle\left\langle\bmath{v}\right\rangle\right\rangle=\bmath{v}/\left|\bmath{v}\right|$.\footnote{In astrometric literature vector normalization is usually denoted as $\left\langle\ldots\right\rangle$. We, however, keep single angular brackets to signify averaging (see Appendix~\ref{app:fitting}) and use double brackets for normalization to avoid confusion.} This equation takes account of proper motion, parallax and perspective (or secular) acceleration. Moreover, the direction $\bmath{u}$ is supposed to be corrected for relevant physical effects such as stellar aberration, gravitational light deflection \citep{klioner2003} and light-travel time \citep{Butkevich+Lindegren2014}.

From here on, by observer's position we mean the barycentric position of the Earth. Thus, we do not differentiate between ground- and space-based observations from near-1AU vantage points; the difference is truly negligible in the context of companion detection.

For a planetary system, its centre of mass is assumed to move uniformly and the direction to the system's host star becomes
\begin{equation}\label{eq:uExo}
 \bmath{u}=
 \left\langle\left\langle\bmath{r}+\left(\bmath{r}\mu_r+\bmath{\mu}\right)\left(t-t_0\right)-
 \varpi\left(\bmath{b}-\bmath{s}\right)/A\right\rangle\right\rangle\,,
\end{equation}
where $\bmath{s}$ is the star position relative to the centre of mass.

The effect of the Earth orbital motion is particularly clear in the special case of a one-planet system with nearly circular orbit when the orbit plane is parallel to the ecliptic. The parallactic ellipse and apparent orbital motion of the host star are then geometrically similar but differing in size. If, moreover, orbital period equals to one year and the star--planet line is in alignment with the Earth position vector, i.e.
\begin{equation}\label{eq:kappa}
 \bmath{s}=k\bmath{b}\,,
\end{equation}
where $k$ is some constant, Eq.~(\ref{eq:uExo}) goes over into the formula (\ref{eq:uStar}) for a single star with the parallax multiplied by the factor $1-k$. Observations are, therefore, fully described by the standard astrometric model with the biased parallax estimate
\begin{equation}
  \varpi_\rmn{est}=\left(1-k\right)\varpi\,.
\end{equation}
This relation shows that, depending of the sign of $k$, the inferred parallax can overestimate or underestimate the true value. This fact is evident from simple considerations. For a parallel alignment, when $k>0$, apparent motion of the host star can be described by a single-star model with a smaller parallax value, whereas for an antiparallel alignment, when $k<0$, a larger parallax value is needed.

\subsection{Parallax bias}
\label{ss:bias}

Equation (\ref{eq:kappa}) represents an exceptional case where the host star and Earth move synchronously. We now consider how correlation between the star and Earth positions affects parallax estimate in general. We make two simplifications to study the parallax absorption of orbital motion in its purest form. First, we assume that the positions and proper motions are known and the parallax is the only parameter to fit. This makes analytical treatment of the problem quite straightforward. Second, we consider the direction to the host star as continuous observable. This enables us to ignore practical aspects of observation and reduction of astrometric data. It is shown in Appendix~\ref{app:fitting} that the $\chi^2$ statistic for the parallax estimation is (cf. Eq.~(\ref{eq:chi2}))
\begin{equation}\label{eq:chi2-replica}
  \chi^2\left(\varpi_\rmn{est}\right)=
  \left\langle
   \left(\varpi\left(\bmath{b}-\bmath{s}\right)/A-\varpi_\rmn{est}\bmath{b}/A\right)^2
  \right\rangle\,.
\end{equation}
where $\varpi_\rmn{est}$ and $\varpi$ are the estimated and true values of the parallax.
Equating to zero the derivative $\rmn{d}\chi^2/\rmn{d}\varpi_\rmn{est}$, we obtain
\begin{equation}\label{eq:profile}
 \left\langle\left[\varpi\left(\bmath{b}-\bmath{s}\right)-\varpi_\rmn{est}\bmath{b}\right]\bmath{\cdot}
 \bmath{b}\right\rangle=0\,.
\end{equation}
This equation yields the parallax estimate
\begin{equation}\label{eq:par-est-def}
 \varpi_\rmn{est}
 =
 \varpi
 \left(1-
 \frac{\left\langle\bmath{s}\bmath{\cdot}\bmath{b}\right\rangle}
 {\left\langle b^2\right\rangle}
 \right)\,.
\end{equation}
Since the eccentricity of the Earth orbit is low, $e=0.017$, we make a very small error assuming that the mean square of the Earth barycentric distance is constant, $\left\langle b^2\right\rangle=A^2$.

Introducing the correlation function of the positions
\begin{equation}\label{eq:corrfunc-def}
  \rho
  \equiv
  \frac{\left\langle\bmath{s\cdot b}\right\rangle}
       {\left\langle b^2\right\rangle^{1/2}\left\langle s^2\right\rangle^{1/2}}
  =
  \frac{\left\langle\bmath{s\cdot b}\right\rangle}
       {A\left\langle s^2\right\rangle^{1/2}}\,,
\end{equation}
we can write Eq.~(\ref{eq:par-est-def}) as
\begin{equation}\label{eq:par-est-corr}
 \varpi_\rmn{est}
 =
 \varpi
 \left(1-\rho
 \frac{\left\langle s^2\right\rangle^{1/2}}{A}\,
 \right)\,.
\end{equation}
The quantity $\rho$ takes on values from $-1$ to $+1$. The relation between $\rho$ and $\left\langle\bmath{s\cdot b}\right\rangle$ is especially simple in the practically important case of circular orbit of radius $a_\star$, where $\left\langle s^2\right\rangle=a_\star^2$, hence
\begin{equation}
  \left\langle\bmath{s\cdot b}\right\rangle=a_\star A\rho
\end{equation}
and Eq.~(\ref{eq:par-est-corr}) for the estimated parallax becomes
\begin{equation}\label{eq:par-est-circ}
 \varpi_\rmn{est}=\varpi\left(1-\frac{a_\star}{A}\,\rho\right)\,.
\end{equation}
This formula gives the complete solution to the problem for a circular orbit: it determines the parallax estimate in terms of the given position correlation and linear size of the star orbit.

We may note that up until now no assumption has been made concerning the mass of the unseen companion, hence Eq.~(\ref{eq:par-est-corr}) is valid of a companion of any mass. For substellar objects, when $a_\star\ll A$, it implies that parallax shift is small compared to the parallax itself. In contrast, presence of a faint companion of stellar mass may result in a large parallax bias.  For a system with an orbital period close to one year, it follows from the Kepler's third law that
\begin{equation}\label{eq:kepler}
  \frac{a_\star}{A}\simeq\left(\frac{M_p}{M_{\sun}}\right)\left(\frac{M_\star}{M_{\sun}}+\frac{M_p}{M_{\sun}}\right)^{-2/3}\,,
\end{equation}
 with $M_\star$ and $M_\rmn{p}$ being the host star and companion mass, respectively. For example, if the host star and companion are both of the solar mass and the orbital period is close to one year, then $a_\star=0.63A$ and $\varpi_\rmn{est}$ ranges from $0.37\varpi$ to $1.63\varpi$ as $\rho$ goes from 1 to $-1$. However, it should be kept in mind that this conclusion is flawed when the companion is sufficiently bright to affect the photocentre motion of the host star.

\subsection{Astrometric signature}
\label{ss:signature}

The angular size of the apparent semi-major axis of the host star orbit, generally referred to as the astrometric signature, is used as a good proxy for assessing astrometric detectability of exoplanets \citep[e.g.,][]{Lattanzi+2000,Casertano+2008,Perryman2011,Perryman+2014,Sozzetti+2014,2017arXiv1704.02493}. If the astrometric signature $\upsilon$ and parallax are measured in arcsec, we can write\footnote{We follow the notation introduced by \cite{2017arXiv1704.02493}; other designations for the astrometric signature also exist in the literature.}
\begin{equation}\label{eq:astr-sig-geom}
  \upsilon\equiv\varpi\frac{a_\star}{A}=
  \left(\frac{M_\rmn{p}}{M_\star}\right)
  \left(\frac{a_\rmn{p}}{A}\right)
  \left(\frac{d}{1\ \rmn{pc}}\right)^{-1},
\end{equation}
where $d$ is the distance, $a_\rmn{p}=\left(M_\star/M_\rmn{p}\right)a_\star$ is semi-major axis of the planet orbit, with $M_\star$ and $M_\rmn{p}$ being the host star and planet mass, respectively; this formula implies $M_\rmn{p}\ll M_\star$.

Defined in such purely geometrical manner, the astrometric signature takes account of neither observer motion nor the fact that exoplanet detection relies on residual analysis. These features, however, can be easily incorporated into the concept of the astrometric signature as shown below.

Astrometric search for exoplanetary systems is based on an analysis of residuals to single star model. However, if orbital motion of the host star is somehow absorbed by the parallax parameter, a specific pattern of orbital motion becomes less pronounced in residuals. If full absorption happens for some system ($\rho=1$), the pattern disappears completely making such system astrometrically undetectable. This fact suggests relating astrometric detectability to an integral characteristic of the parallax fit residuals. The residuals are conveniently quantified by the minimum value of the $\chi^2$ statistic, $\chi_\rmn{min}^2$, attained with the estimated parallax. It is demonstrated below that, when no parallax absorption occurs, the square root of $\chi_\rmn{min}^2$ coincides with the conventional astrometric signature given by Eq.~(\ref{eq:astr-sig-geom}). At the same time, it describes the corresponding effect on the residuals if parallax is fully or partially absorbed. We, therefore, refer to this quantity as the effective astrometric signature, $\upsilon_\rmn{eff}$, hereafter. Substituting Eq.~(\ref{eq:par-est-def}) into Eq.~(\ref{eq:chi2-replica}), we find after some simple algebra the minimum of $\chi^2$
\begin{equation}\label{eq:chi2min}
  \chi_\rmn{min}^2\equiv\chi^2\left(\varpi_\rmn{est}\right)=
  \left(1-\rho^2\right)
  \frac{\varpi^2}{A^2}\left\langle s^2\right\rangle\,;
\end{equation}
this gives the effective astrometric signature
\begin{equation}\label{eq:astr-sig-def}
  \upsilon_\rmn{eff}\equiv
  \left(\chi_\rmn{min}^2\right)^{1/2}=
  \left(1-\rho^2\right)^{1/2}\frac{\varpi}{A}
  \left\langle s^2\right\rangle^{1/2}\,.
\end{equation}
In the case of a circular orbit, we obtain the following expression, which differs from the conventional definition of the astrometric signature only by a factor depending on the correlation function:
\begin{equation}\label{eq:astr-sig-eff}
  \upsilon_\rmn{eff}
  =\left(1-\rho^2\right)^{1/2}\varpi\frac{a_\star}{A}
  \equiv\kappa\upsilon\,.
\end{equation}
This formula generalizes Eq.~(\ref{eq:astr-sig-geom}) to the case of moving observer, including residual effects. We now consider the new quantity
\begin{equation}\label{eq:susceptibility}
  \kappa=\left(1-\rho^2\right)^{1/2}\,.
\end{equation}
Since $\left|\rho\right|\lid 1$, $\kappa$ runs from 0 to 1. $\kappa=0$ corresponds to complete absorption of the host star orbital motion by the parallax parameter, and $\kappa=1$ for uncorrelated positions when the host star orbital motion has no effect on parallax estimation. Accordingly, no signature of orbital motion remains in the parallax fit residuals when $\kappa=0$, whereas for $\kappa=1$ it completely propagates into the residual. Thus the parameter $\kappa$ is a measure for the fraction of astrometric information on a given system which goes into residual analysis in the context of orbital motion recognition. For this reason, the quantity $\kappa$ is referred to as the \emph{astrometric efficiency} in what follows. In addition to the $\rho$, $\kappa$ is also a convenient measure for astrometric detectability of exoplanetary systems.

\subsection{Relation between parallax bias and astrometric signature}

Formulae derived in the preceding sections reveal a remarkable relationship between the parallax bias and astrometric signature. Namely, comparison of Eqs.~(\ref{eq:par-est-circ}) and (\ref{eq:astr-sig-geom}) gives
\begin{equation}\label{eq:parBias-astrSignature}
  \Delta\varpi=-\rho\upsilon\,,
\end{equation}
where $\Delta\varpi=\varpi_\rmn{est}-\varpi$ is the parallax bias. It should be emphasized that this equation is valid for a parallax bias due to star orbital motion; however, it does not hold in general for a bias originating from other reasons.

Using this relation, one can calculate any of the quantities $\Delta\varpi$, $\rho$ and $\upsilon$ provided that the other two are known. For instance, if information on $\rho$ and $\upsilon$ is gained from radial velocity or transit observations, it in principle may serve to constrain the parallax bias. Moreover, this relation enables us to draw several qualitative conclusions. First, if the correlation function or even only its sign is known, a sign of the parallax bias can be predicted. Second, since $\rho$ lies between $-1$ and $+1$, absolute value of the parallax bias cannot exceed the astrometric signature, $\left|\Delta\varpi\right|\lid\upsilon$. These theoretical predictions deserve further experimental verification in numerical simulations.

It is useful to notice one further relation, which does not explicitly contains the correlation function and involves the parallax shift, astrometric signature and minimum value of $\chi^2$. Squaring Eq.~(\ref{eq:parBias-astrSignature}) and making use of Eq.~(\ref{eq:chi2min}) for a circular orbit, we obtain
\begin{equation}
  \left(\Delta\varpi\right)^2+\chi_\rmn{min}^2=\upsilon^2\,.
\end{equation}
This relation reflects information balance between parallax estimation and astrometric planet detection. To avoid misunderstanding, it is worth mentioning that we are here only concerned with the parallax shift owing to orbital motion of the host star.

\section{Correlation function}\label{s:application}

\begin{figure}
  \centering
  \includegraphics[width=0.875\hsize]{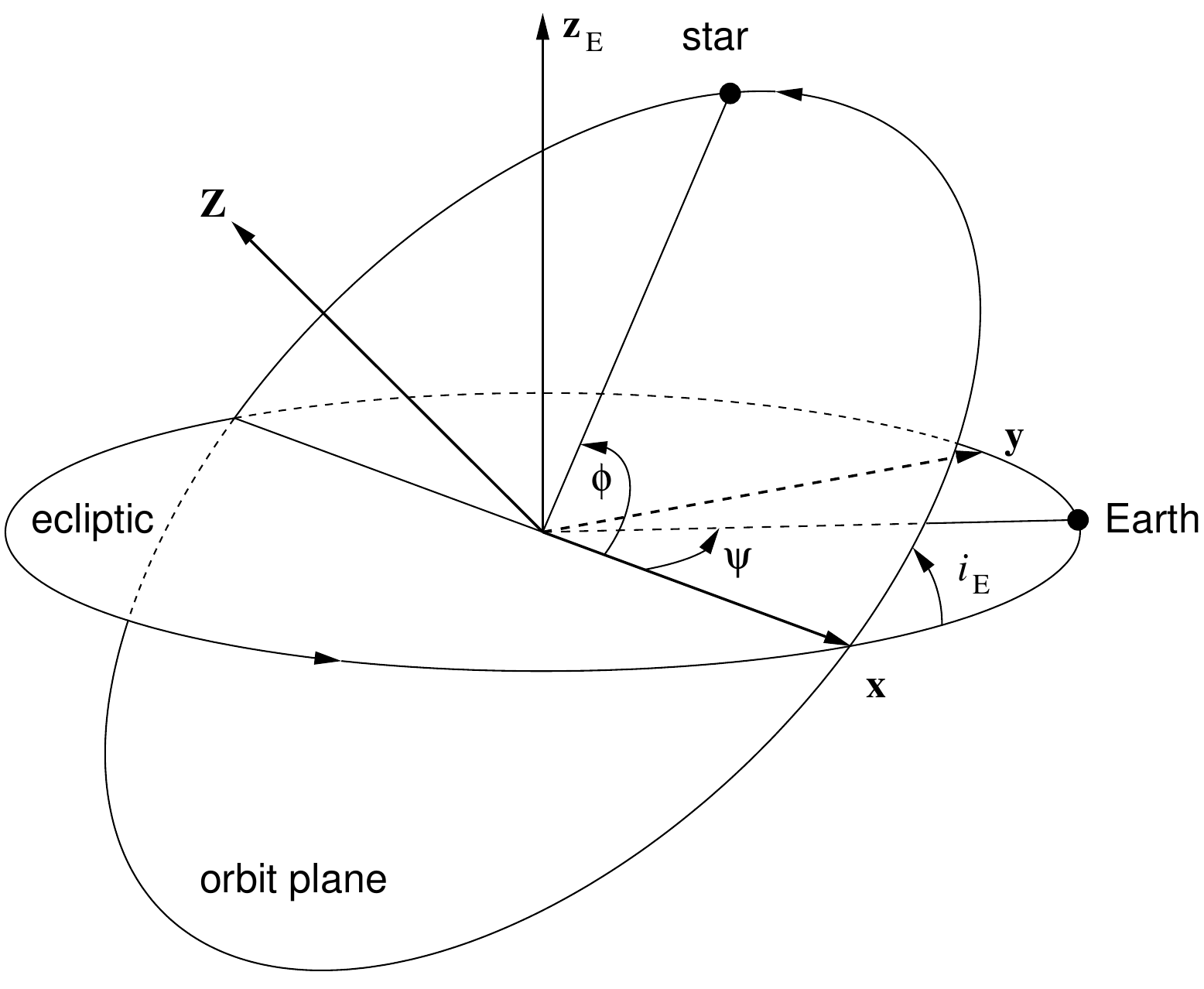}
 \caption{Orientation of the circular orbit with respect to the reference plane parallel to the ecliptic. Projection of the initial position of the Earth on to the reference plane is also shown. The orbit sizes are not in scale. The arrows indicate directions of the host star orbital motion induced by the unseen companion and the Earth's revolution around the solar system barycentre. The inclination of the orbital plane to the ecliptic $i_\rmn{E}$ is given by Eq. (\ref{eq:cosie}). The reference frame is defined by the ecliptic polar axis $\bmath{z}_\rmn{E}$ and the line of nodes, along which the orbital and reference planes intersect. The positive direction of the $x$ axis is defined by the ascending node where the star crosses the reference plane in the positive $z$ direction. The initial orbital phases $\phi$ and $\psi$ are the positions of the star and Earth, respectively, at $t=0$ measured from the $x$ axis. $\bmath{Z}$ is the unit vector along the host star's angular momentum around its system barycentre.}
  \label{fig:orbit}
\end{figure}

In this section we consider basic properties of the correlation function. We first introduce parameters to describe orientation of the host orbit with respect to the ecliptic as well as orbital positions of the star and Earth. Using this parametrization, we then calculate the correlation function for a circular orbit. Finally, we derive, under some simplifying assumptions, thej dependence of the correlation function on eccentricity for an elliptical orbit.

\subsection{Orbit orientation with respect to the ecliptic}
\label{ss:orientation}

In the equatorial and ecliptic systems, the polar axes are represented by the unit vectors $\bmath{z}$ and $\bmath{z}_\rmn{E}$ towards $\delta=+90\degr$ and $\beta=+90\degr$. At a position on the celestial sphere, determined by the reference direction $\bmath{r}$, let the orthogonal unit vectors $\bmath{p}$, $\bmath{q}$ and $\bmath{p}_\rmn{E}$, $\bmath{q}_\rmn{E}$ specify the directions of increasing $\left(\alpha,\delta\right)$ and $\left(\lambda,\beta\right)$, respectively. The vector pairs  $\bmath{p}$, $\bmath{q}$ and $\bmath{p}_\rmn{E}$, $\bmath{q}_\rmn{E}$ are the tangential constituents of the so-called normal triads relative to the equatorial and ecliptic systems. Their coordinate representations are given in Appendix~\ref{app:angle}.

Since, as we have seen, the correlation between the host star and Earth positions depends crucially on mutual orbit orientation, it is useful to describe it in terms of the orbit inclination to the ecliptic, $i_\rmn{E}$, defined as the angle between $\bmath{z}_\rmn{E}$ and the star's angular momentum vector (see Fig.~\ref{fig:orbit}). The orbital motion is direct, i.e. the host star and Earth revolve in the same direction, if $i_\rmn{E}<\upi/2$, while the star proceeds in a reverse direction if $i_\rmn{E}>\upi/2$.

The quantity $i_\rmn{E}$ can be found as follows. We introduce the position angle $\theta$ of the vector $\bmath{p}_\rmn{E}$, i.e. the angle between $\bmath{q}$ and $\bmath{p}_\rmn{E}$ reckoned counterclockwise from $\bmath{q}$. For reasons that are clearified below, it is convenient to set its range from $-\upi$ to $\upi$. The angle $\theta$ is a measure for mutual orientation of the equatorial and ecliptic coordinate grids. The calculation of $\theta$ is described in Appendix~\ref{app:angle}. Using Eq.~(\ref{eq:pq-pairs}),
the unit vector towards the north ecliptic pole can be written as
\begin{equation}
    \bmath{z}_\rmn{E}=
    \bmath{q}_\rmn{E}\cos\beta+\bmath{r}\sin\beta=
    \left(-\bmath{p}\cos\theta+\bmath{q}\sin\theta\right)\cos\beta+\bmath{r}\sin\beta\,.
\end{equation}

Let $\bmath{Z}$ be the unit vector along the host star's angular momentum. It is normal to the orbit plane and expressed in terms of the standard Keplerian elements as
\begin{equation}
 \begin{aligned}
    \bmath{Z}
    &=
    \bmath{p}\sin i\cos\Omega-
    \bmath{q}\sin i\sin\Omega-
    \bmath{r}\cos i\\
    &=
    -\bmath{p}_\rmn{E}\sin i\cos\left(\Omega-\theta\right)
    -\bmath{q}_\rmn{E}\sin i\sin\left(\Omega-\theta\right)
    -\bmath{r}\cos i\,,
 \end{aligned}
\end{equation}
where $i$ is the orbit inclination to the tangent plane and $\Omega$ is the position angle of ascending node.

Taking the dot product of $\bmath{z}_\rmn{E}$ and $\bmath{Z}$, we finally obtain
\begin{equation}\label{eq:cosie}
 \cos i_\rmn{E}=\bmath{Z\cdot z}_\rmn{E}=
 -\sin\beta\cos i-\cos\beta\sin i\cos\left(\Omega-\theta\right)\,.
\end{equation}
This formula gives the solution to the problem: it determines $i_\rmn{E}$: the inclination of the circular orbital with respect to the ecliptic in terms of the coordinates and orbital elements.

The orbit inclination $i$, which is one of seven Keplerian elements in standard orbit specification, is not of our main interest here. Instead we are interested in $i_\rmn{E}$. For brevity, we therefore call the quantity $i_\rmn{E}$ simply the inclination from here on and only explicitly mention that it refers to the ecliptic where it is needed to avoid ambiguity.

We now proceed to find out in what cases $i_\rmn{E}$ is zero, that is, when the orbit plane is parallel to the ecliptic and the orbital motion is direct. Equation~(\ref{eq:cosie}) in this case gives
\begin{equation}\label{eq:cosie0}
  \sin\beta\cos i+\cos\beta\sin i\cos\left(\Omega-\theta\right)=-1\,.
\end{equation}
After simple transformations, this formula can be written as
\begin{equation}
  \cos\beta\sin i\left(\cos\left(\Omega-\theta\right)+1\right)=\sin\left(i-\beta\right)-1\,.
\end{equation}
It is evident that the right-hand side never exceeds zero, whereas the left-hand side is non-negative for $-\upi/2\lid\beta\lid\upi/2$ and $0\lid i\lid\upi$. Hence the equality holds only if both the sides are equal to zero; that is if $\cos\left(\Omega-\theta\right)=-1$ and $\sin\left(i-\beta\right)=1$. Thus, $i_\rmn{E}=0$ when
\begin{align}\label{eq:condition-i-beta}
  i&=\beta+\upi/2\,,\\
     \label{eq:condition-Omega-theta}
  \Omega&=\theta+\upi\,.
\end{align}
These conditions lend themselves to a straightforward geometrical interpretation.

\subsection{Circular orbit}\label{ss:correlation}

The reference frame is defined by the vectors $\bmath{z}_\rmn{E}$ and $\bmath{Z}$, with the positive $x$ direction along $\bmath{z}_\rmn{E}\bmath{\times}\bmath{Z}$ and  $\bmath{y}=\bmath{z}_\rmn{E}\bmath{\times}\bmath{x}$, as shown in Fig.~\ref{fig:orbit}. Let $\phi$ and $\psi$ be the angular distances of the star and Earth, respectively, from the $x$ axis at the moment of the beginning of observations $t=0$. The positions of Earth and the host star are then
\begin{align}
 \bmath{b}&=
 \begin{pmatrix}
    A\cos\Psi, &
    A\sin\Psi, &
    0
 \end{pmatrix}\,,\label{eq:b}\\
 \bmath{s}&=
 \begin{pmatrix}
    a_\star\cos\Phi, &
    a_\star\sin\Phi\cos i_\rmn{E}, &
    a_\star\sin\Phi\sin i_\rmn{E}
 \end{pmatrix}\,.\label{eq:s}
\end{align}
Here $\Psi=\psi+2\upi\,t$ and $\Phi=\phi+2\upi\,t/P$ are the orbital phases, with the time $t$ and the star orbital period $P$ measured in Julian years.

Substituting Eqs.~(\ref{eq:b}) and (\ref{eq:s}) into Eq.~(\ref{eq:corrfunc-def}) and employing Eq.~(\ref{eq:time-average}) for the averages yield the correlation function
\begin{equation}\label{eq:corrfunc}
 \begin{aligned}
  \rho =
  \frac{\sin\left[\upi\,T\left(1-1/P\right)\right]}{\upi\,T\left(1-1/P\right)}
  &\cos\left[\upi\,T\left(1-\frac{1}{P}\right)+\psi-\phi\right]
  \cos^2\frac{i_\rmn{E}}{2} \\
  +
  \frac{\sin\left[\upi\,T\left(1+1/P\right)\right]}{\upi\,T\left(1+1/P\right)}
  &\cos\left[\upi\,T\left(1+\frac{1}{P}\right)+\psi+\phi\right]
  \sin^2\frac{i_\rmn{E}}{2}\,,
 \end{aligned}
\end{equation}
where $T$ is the duration of observations in years. In the derivation of Eq.~(\ref{eq:corrfunc}) no assumption has been made concerning the period value or observation length, it is, therefore, formally valid for any $P$ and $T$. From a practical viewpoint, however, this equation is meaningful only if the observations span at least one orbital period of the host star. In the special case of $P=1$~yr, Eq.~(\ref{eq:corrfunc}) reduces to
\begin{equation}\label{eq:corrfunc1yr}
 \begin{aligned}
  \rho =
  &\cos\left(\psi-\phi\right)
  \cos^2\frac{i_\rmn{E}}{2} \\
  +
  &\frac{\sin\left(2\upi\,T\right)}{2\upi\,T}
  \cos\left(2\upi\,T+\psi+\phi\right)
  \sin^2\frac{i_\rmn{E}}{2}\,.
 \end{aligned}
\end{equation}
It is easily seen that for large $T$ this expression tends toward the constant value
\begin{equation}\label{eq:corrfunc-appr}
  \rho=\cos\Delta\,\cos^2\frac{i_\rmn{E}}{2}\,,
\end{equation}
where $\Delta=\psi-\phi$ is the initial phase shift.
Thus the limiting value of the correlation function depend on two parameters, the inclination and phase difference at $t=0$. Furthermore, we note that the simple formula (\ref{eq:corrfunc-appr}) is very useful for analytical estimations because it provides a good approximation to the exact expression (\ref{eq:corrfunc1yr}) for $T\ga 1$~yr.

If $P=1$~yr and the orbit plane is parallel to the ecliptic and the orbital motion is direct, i.e. $i_\rmn{E}=0$,
\begin{equation}
  \rho = \cos\Delta
\end{equation}
and Eq.~(\ref{eq:susceptibility}) gives the astrometric efficiency
\begin{equation}
  \kappa = \left|\sin\Delta\right|\,.
\end{equation}
Thus, both $\rho$ and $\kappa$ depend only on the absolute value of the orbital phase difference in this case. The correlation function $\rho$ runs from $+1$ for parallel $\left(\Delta=0\right)$ to $-1$ for antiparallel $\left(\Delta=\pm\upi\right)$ alignment, passing through zero for the orthogonal configuration $\left(\Delta=\pm\upi/2\right)$. The astrometric efficiency $\kappa$ accordingly attains its maximum value of 1 for orthogonal geometry and vanishes both for parallel and antiparallel alignments.

\subsection{Elliptical orbit: simplified treatment}
\label{ss:ellipse}

Calculation of the correlation function in the general case of elliptical orbit presents great mathematical difficulties. However, in the context of the present work it is sufficient to study the correlation function only in a very narrow range in the parameter space where it is close to unity. The calculations are relatively simple under the following assumptions.

First, we assume that the orbital plane is parallel to the ecliptic, $i_\rmn{E}=0$, and choose the axis $x$ along the semimajor axis towards the pericentre. Second, let the positions of the host star and Earth be in parallel alignment at $t=0$, i.e. the initial orbital phases $\phi$ and $\psi$ be zero. It is worth noting that the condition $\phi=0$ is equivalent to the statement that the host star passes pericentre of its orbit at the moment of the beginning of observations. Third, we consider a system with $P=1\,\rmn{yr}$ and suppose that the observations cover the entire orbital period. This allows us to describe the results compactly in terms of the Bessel functions of integral order, $J_n$.

Appendix~\ref{app:ellipse} gives calculation of the mean values $\left\langle s^2\right\rangle$ and $\left\langle\bmath{s\cdot b}\vphantom{s^2}\right\rangle$ based on these assumptions.
Combining Eqs.~(\ref{eq:app-ell-s2}) and (\ref{eq:app-ell-sb}), we find that the dependence of the correlation function on the eccentricity has the form
\begin{equation}\label{eq:ell-corr}
  \rho=\left(1+\frac{3}{2}e^2\right)^{-1/2}\left(J_0\left(e\right)-\frac{1-\sqrt{1-e^2}}{e}J_1\left(e\right)\right)\,.
\end{equation}
This dependence is illustrated in Fig.~\ref{fig:elliptic}.

Equation~(\ref{eq:app-ell-sb-approx}) shows that for small eccentricity this formula is approximated by
\begin{equation}\label{eq:ell-corr-approx}
  \rho=1-\frac{5}{4}e^2\,.
\end{equation}
Substitution of Eq.~(\ref{eq:ell-corr-approx}) into (\ref{eq:susceptibility}) gives astrometric efficiency
\begin{equation}
  \kappa=e\sqrt{5/2}\,.
\end{equation}
Thus, if host star's orbit slightly deviates from a circle, the astrometric efficiency increases linearly with eccentricity, while the correlation function and parallax bias are quadratic functions of eccentricity. In the opposite extreme case of very elongated orbit, which formally corresponds to the limit of $e\to 1$, the correlation function approaches its minimum value of $\sqrt{2/5}\left(J_0\left(1\right)-J_1\left(1\right)\right)=0.21$; accordingly the efficiency attains the maximum value of 0.98. We address the effect of eccentricity on astrometric detectability further in Section~\ref{ss:eff-ellipse}.

\section{Discussion}\label{s:discussion}

In the preceding sections, we showed that absorption of orbital motion by the parallax parameter affects astrometric detectability of exoplanetary systems. Here we proceed to discuss some aspects of this problem. We consider first the impact of the orbital phases and period on the detectability. Since detection is very sensitive to system parameters when the period is close to one year, we examine such systems further.
Starting from the effects of eccentricity and inclination, we consider then limit of long observations, and finally estimate the accompanying decrease in the detection probability. Furthermore, we briefly review advantages of employing other detection techniques.

\subsection{Effects of orbital phases and period}
\label{ss:phase}

\begin{figure}
  \centering
  \includegraphics[width=\hsize]{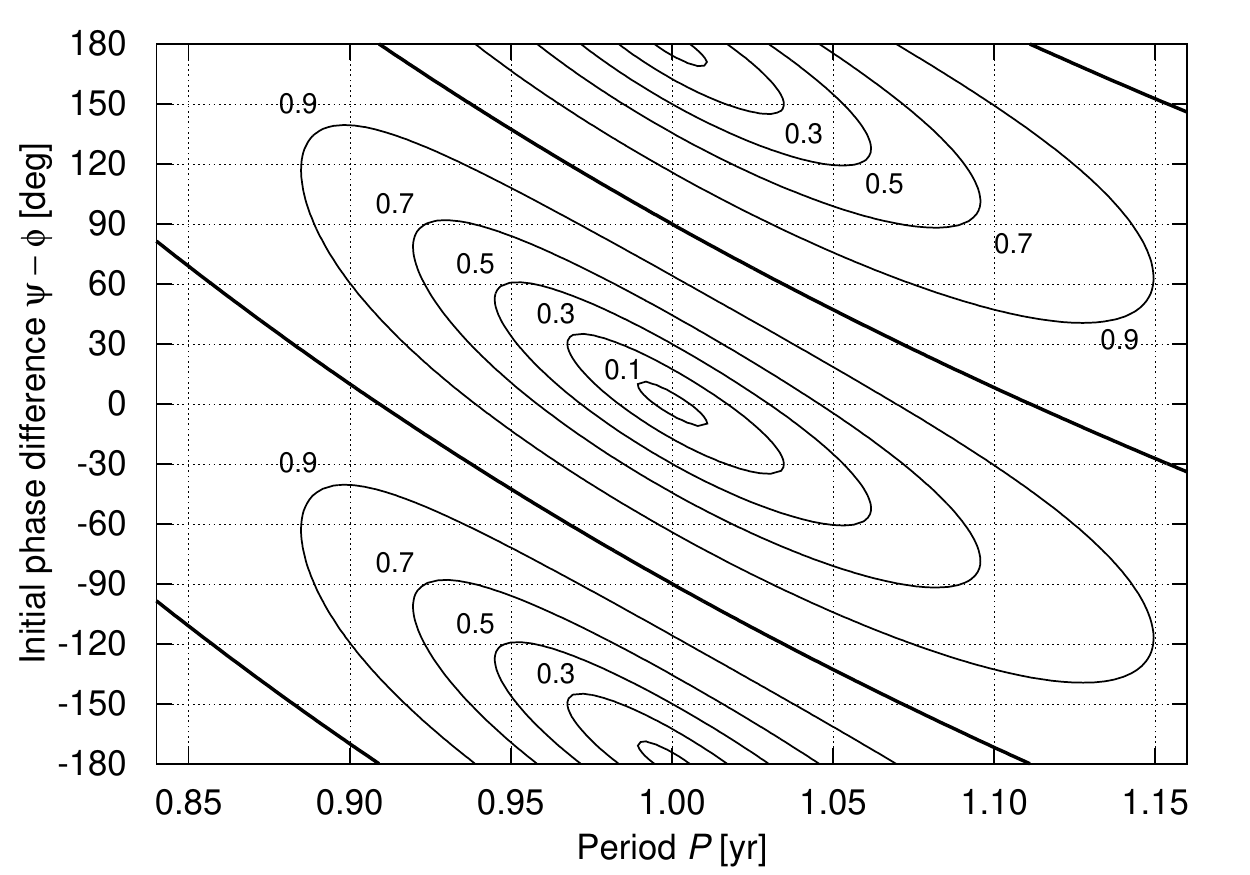}
 \caption{Contour plot of the astrometric efficiency $\kappa$ as a function of the orbital period $P$ and initial phase difference $\Delta=\psi-\phi$. The labels indicate values of $\kappa$ for the isolines calculated from Eqs.~(\ref{eq:susceptibility}) and (\ref{eq:corrfunc}) with duration of observations $T=5$~yr and inclination $i_\rmn{E}=0\degr$ (i.e. for a direct coplanar motion of the host star). The heavy lines show where $\kappa=1$ and the orbital motion of the host star is not absorbed by the parallax estimate. The zero values attained at $P=1$~yr and the phase differences of 0\degr and $\pm 180\degr$ correspond to parallel and anti-parallel alignment of the Earth and star position vectors, for which the entire orbital motion of the host star is absorbed by the parallax estimate.}
 \label{fig:phase}
\end{figure}

For simplicity, we discuss here a circular coplanar orbit; arbitrary inclinations are further addressed in Sect.~\ref{ss:long-observations}. Equation~(\ref{eq:corrfunc}) suggests that for $i_\rmn{E}=0$ the correlation function $\rho$ depends solely on the initial orbital phase shift $\Delta$. For illustrative purposes it is convenient to consider in place of $\rho$ the astrometric efficiency $\kappa$, given by Eq.~(\ref{eq:susceptibility}). As explained in Sect.~\ref{ss:signature}, this quantity is very useful is studying fine details of detection probability. Figure~\ref{fig:phase} exemplifies behaviour of $\kappa$ for $T=5$~yr. Inspection of this plot shows that $\kappa$ is very sensitive to the phase shift if the period is close to one year and becomes weakly dependent on it otherwise.

This dependence on the period results from the fact that the position vectors cease rotating synchronously and correlation between them drops rapidly as the period deviates from one year. Numerical calculations indicated that, for given $\kappa$ and $T$, the detection probability crucially depends on the orbital phases for $1-\Delta P<P<1+\Delta P$, where $\Delta P=\left(0.76\kappa-0.08\right)/T$ for a coplanar orbit. Inclination slightly narrows the interval $\Delta P$. Outside this period range effect of the phases is practically negligible.

\subsection{Dependence on survey duration. Decorrelation time}
\label{ss:long-observations}

\begin{figure}
  \centering
  \includegraphics[width=\hsize]{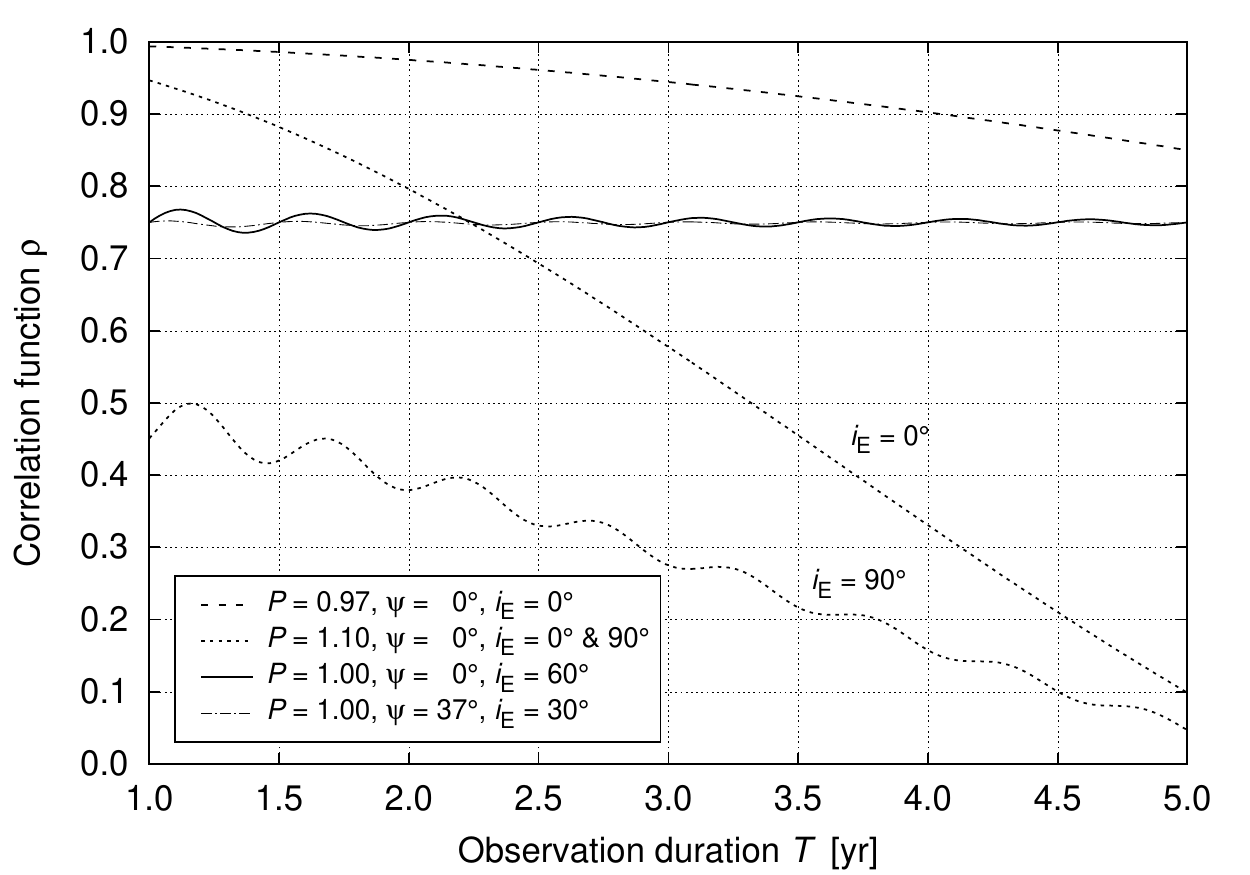}
 \caption{Dependence of the correlation function $\rho$ on duration of observations for $\phi=0\degr$. The dotted lines show $\rho$ calculated from Eq.~(\ref{eq:corrfunc}) with $P=1.1$~yr and $\psi=0\degr$ for two different inclinations, 0\degr and 90\degr, indicated next to the curves, while the dashed curve assumes a coplanar orbit with $P=0.97$~yr and the same $\psi$. The other graphs illustrate the case of $P=1$~yr and show $\rho$ calculated from Eq.~(\ref{eq:corrfunc1yr}). The results for $i_\rmn{E}=60\degr$, $\psi=0\degr$ and $i_\rmn{E}=30\degr$, $\psi=37\degr$ are drawn with the solid and dash-dotted lines.}
  \label{fig:effdurat}
\end{figure}

The behaviour of the correlation function when observations cover a long time span depends considerably on orbital period. Plots of $\rho$ versus $T$ for several sets of parameters are shown in Fig~\ref{fig:effdurat}. If the period is not equal to one year, $\rho$ decreases with increasing $T$ and vanishes in the long run. It means that orbital motion does not affect detectability provided that observations are long enough. The reason for this is the lack of synchronism between the orbital motions of the Earth and host star discussed in Sect.~\ref{ss:phase}.

The time within which the correlation function vanishes is called the decorrelation time $T_\rmn{d}$. Analysis of Eq.~(\ref{eq:corrfunc}) shows that, to order of magnitude,
\begin{equation}
 T_\rmn{d}\sim\frac{P}{\left|P-1\right|}\,.
\end{equation}
In using the term "sufficiently long" intervals of time, we have meant essentially times long compared with the decorrelation time. For example, for the the cases of $P=1.1$ and 0.97~yr shown in Fig.~\ref{fig:effdurat}, $T_\rmn{d}=11$ and 32~yr, respectively. Thus it can take quite a long time for correlation to vanish.

For a period strictly equal to one year the situation is different. As demonstrated in Sect.~\ref{ss:correlation}, in this case the correlation function tends toward the constant value given by Eq.~(\ref{eq:corrfunc-appr}).

\subsection{Effects of eccentricity and inclination}
\label{ss:eff-ellipse}

\begin{figure}
  \centering
  \includegraphics[width=\hsize]{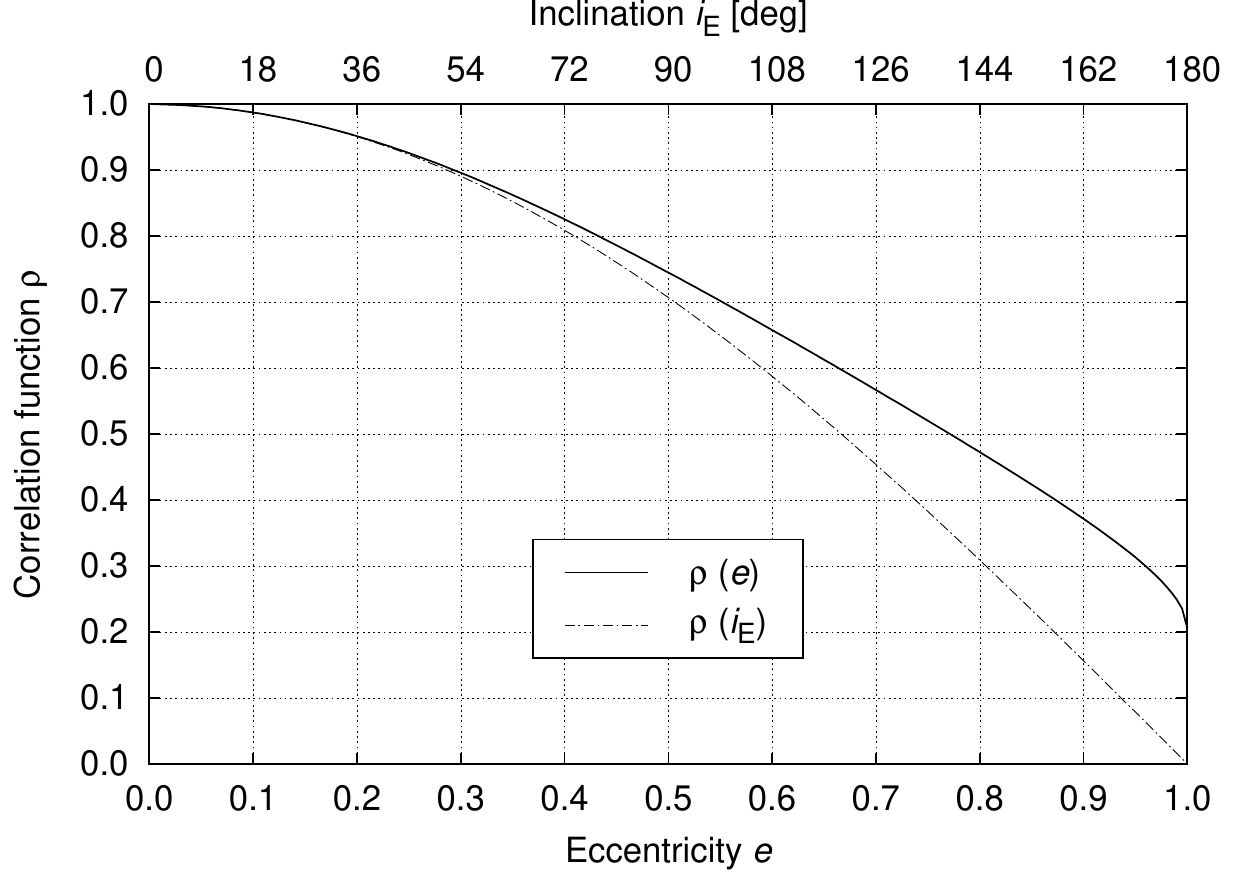}
 \caption{Dependence of the correlation function on eccentricity and orbit inclination for $P=1$~yr. The solid line and lower axis shows $\rho$ as a function $e$ calculated from Eq.~(\ref{eq:ell-corr}), while the dash-dotted line and upper axis show $\rho$ as a function of $i_\rmn{E}$ calculated from Eq.~(\ref{eq:corrfunc1yr}) for $\Delta\left(\equiv\psi-\phi\right)=0$.}
  \label{fig:elliptic}
\end{figure}

To illustrate effects of ellipticity and projection on astrometric detectability, the plots of $\rho$ versus $e$ and $i_\rmn{E}$ are shown in Fig.~\ref{fig:elliptic}. The graphs are intentionally drawn for the simplest case of $P=1$~yr and $\Delta\left(\equiv\psi-\phi\right)=0$ to clearly demonstrate similarities and differences between circular and elliptic orbits.

For a non-zero inclination, a circular orbit projects into an ellipse\footnote{It is worth recalling that projection onto the ecliptic plane is meant here.} and becomes geometrically similar to an actual elliptical orbit. This resemblance explains why the functions $\rho\left(e\right)$ and $\rho\left(i_\rmn{E}\right)$ coincide for small and moderate values of the parameters, $e\loa0.4$ and $i_\rmn{E}\loa70\degr$. The difference between the functions for the large parameter values is due to the fact that these parameter ranges describe different orbit types. While $e\simeq1$ corresponds to a highly elongated orbit, $i_\rmn{E}$ close to 180\degr corresponds to a retrograde motion on a circular orbit nearly parallel to the ecliptic (see Sect.~\ref{ss:orientation}).

The correlation drops as an orbit becomes more elongated. This decrease in $\rho$ results from the fact that both length and angular velocity of the star position vector are variable for such an orbit, whereas these quantities are constant in the Earth motion. The higher eccentricity is, the larger the difference and the smaller the correlation. This explanation is also applicable to the behaviour of the correlation function for a circular orbit with $i_\rmn{E}<90\degr$ because higher values of the inclination correspond to more elongated projections. However, for $i_\rmn{E}>90\degr$ the correlation decreases since the star and Earth revolve in opposite directions.

\subsection{Effect on detection probability}

Decrease in the astrometric signature due to absorption of orbital motion by the parallax parameter reduces the signal-to-noise ratio $S/N\equiv\upsilon/\sigma$, with $\sigma$ being the astrometric accuracy. By analogy with the effective astrometric signature, we can introduce the effective signal-to-noise ratio:
\begin{equation}
  \left(\rmn{S}/\rmn{N}\right)_\rmn{eff}=\kappa\left(\rmn{S}/\rmn{N}\right)\,.
\end{equation}
If we consider a set of system all having the same S/N (i.e. systems with $\upsilon/\sigma=a_\star/\left(\sigma d\right)=\rmn{const}$), a mean value of the astrometric efficiency may serve as a convenient measure of reduction in detection probability.

As explained in Sect.~\ref{ss:phase}, $\kappa$ is close to zero in a narrow interval of periods around one year. Therefore, we can simply put $P=1$~yr when calculating the mean value. Moreover, we consider duration of observations long enough for the approximate Eq.~(\ref{eq:corrfunc-appr}) to be valid; this gives approximate expression for $\kappa$:
\begin{equation}
  \kappa=\left(1-\cos^2\Delta\,\cos^4\frac{i_\rmn{E}}{2}\right)^{1/2}\,.
\end{equation}
We also assume that the phase shift $\Delta\left(\equiv\psi-\phi\right)$ and orbit orientation $i_\rmn{E}$ are distributed at random (the latter is equivalent to a uniform distribution of $\cos i_\rmn{E}$). The averaging is then done using a formula from the theory of elliptic integrals \citep[see][Section 6.148]{gradshteyn} as follows
\begin{equation}
 \begin{aligned}
  \overline{\kappa}
  &=\frac{1}{4\upi}
    \int\limits_{-1}^1\int\limits_0^{2\upi}
    \left(1-\cos^2\Delta\,\cos^4\frac{i_\rmn{E}}{2}\right)^{1/2}\,
     \rmn{d}\Delta\,\,\rmn{d}\cos i_\rmn{E}\\
  &=\frac{1}{\upi}\int\limits_{-1}^1 E\left(\cos^2\frac{i_\rmn{E}}{2}\right)\rmn{d}\cos i_\rmn{E}
  =\frac{2}{\upi}\int\limits_0^1 E\left(k\right)\rmn{d}k
  =\frac{2G+1}{\upi}\,,
 \end{aligned}
\end{equation}
where $E$ is the complete elliptic integral of the second kind,
\begin{equation}
  E\left(k\right)=\int_{0}^{\upi/2}\left(1-k^2\sin^2\varphi\right)^{1/2}\,\rmn{d}\varphi\,,
\end{equation}
and $G=0.91597...$ is the Catalan's constant.

Numerical value $\overline{\kappa}\simeq0.90$ suggests that, for a population of systems with the period close to one year, S/N might be expected to drop by 10 per cent on average. Number of detections is determined statistically as number of systems for which S/N is above a specified detection threshold. For example, the threshold of $\rmn{S}/\rmn{N}>3$ is often used as rule of thumb in studies on astrometric planet detection \citep{Perryman+2014,Sozzetti+2014,2017arXiv1704.02493}. The 10\% diminution in S/N means that expected number of detected systems with $P\simeq1~\rmn{yr}$ corresponds to the number of systems with the same astrometric signature and period different from one year which would be detected if the threshold were 10 per cent lower.

\subsection{Other detection techniques }

It is worth briefly considering potential of other major detection techniques, photometric transit searches and radial velocity surveys \citep{Perryman2011}, for discovering planetary systems which are astrometrically undetectable because of the absorption of orbital motion by the parallax parameter. We also discuss the possibility of using photometric data to recognize the accompanying parallax bias.

\subsubsection{Transits}

Transit can occur if $\cos i\lid R_\star/a_\rmn{p}$, with $R_\star$ being the radius of the host star. For $M_\rmn{p}\ll M_\star$ and $P\simeq1\,\rmn{yr}$, Kepler's third law implies that $\left(a_\rmn{p}/1\,\rmn{AU}\right)\simeq\left(M_\star/M_{\sun}\right)^{1/3}$. We consider a direct motion of the host star in an orbit plane nearly coplanar to the ecliptic, i.e. $i_\rmn{E}\simeq0$. Using Eq.~(\ref{eq:condition-i-beta}), we can then write the condition for transit as
\begin{equation}
  \left|\beta\right|\loa 0\fdg27\left(\frac{R_\star}{R_{\sun}}\right)\left(\frac{M_\star}{M_{\sun}}\right)^{-1/3}.
\end{equation}
This formula suggests that the systems of interest can manifest themselves as transiting planets only within a narrow strip around the ecliptic. For main sequence stars, $R_\star$ is roughly proportional to $M_\star^{0.8}$ \citep[][Ch.~15]{Cox2000}, hence $\left|\beta\right|\loa0\fdg27\left(M_{\sun}/M_\star\right)^{0.47}$. Accordingly,
the width of the strip ranges from 0\fdg4 for M0 dwarf host stars to about 0\fdg7 for F0V stars.

The proximity of host star to the ecliptic places constraints on observability of planetary transits for coplanar systems. Simple geometric arguments suggest that the host star is separated by $180\degr-\left|\Delta\right|$ from the Sun when the planet moves in front of the star. If the orbital phase shift $\Delta$ is close to $\pm180\degr$, i.e. the position vectors $\bmath{b}$ and $\bmath{s}$ are nearly antiparallel, transits always occur when the host star is either close to the solar limb or even behind the Sun as seen from the Earth. The transits are, therefore, unobservable and secondary eclipses remain the only photometric evidences for the planetary companion. Because of observational limitations, transit event may remain undiscovered even if a configuration significantly deviates from a precise antiparallel alignment. For instance, if a transit search programme is capable of monitoring stars beyond a minimum angular distance $\xi_\rmn{min}$ from the Sun, detection of transits is only feasible for systems with $\left|\Delta\right|<\Delta_\rmn{max}\left(\equiv180\degr-\xi_\rmn{min}\right)$ (we recall that we consider nearly coplanar systems in the vicinity of the ecliptic). Thus, for systems of this kind, there is a selection effect against detecting transits for $\left|\Delta\right|>\Delta_\rmn{max}$.

\subsubsection{Radial velocities}

Radial velocity surveys seem to offer better prospects for discovering astrometrically undetectable exoplanets.
Both astrometric and Doppler observations are corrected for the orbital motion of the Earth.
However, there is a marked difference in the way the corrections are applied.
The standard astrometric model (see Eq.~(\ref{eq:uStar})), with the parallax term accounting for the Earth motion, relies on parallax -- a quantity which either to be determined or subject to uncertainty. If distance to the target system were precisely known, orbital motion of the host star could, in principle, be recovered from astrometric data.

In contrast, the Earth orbital velocity, which enters into the correction of Doppler observations \citep{Wright+Eastman2014}, is very well known. As a result, velocity of the star with respect to the solar system barycentre is accurately determined and its orbital motion can be detected. It is worth mentioning that ,  besides evident limitations imposed by observational accuracies, systems close the ecliptic poles escape detection because of decrease in radial velocity amplitude at high ecliptic latitudes for orbits coplanar to the ecliptic.

\subsubsection{Bias in absolute magnitude}

The parallax shift results in a biased estimation of the host star absolute magnitude. Accordingly, presence of unseen companion can be inferred from difference between trigonometric and photometric or spectroscopic parallaxes. This is obvious for stellar companions when the shift in trigonometric parallax may be comparable to, or even larger, the parallax itself. For example, in the case of a solar mass companion, considered at the end of Sect.~\ref{ss:bias}, the difference between the derived and true absolute magnitude ranges from $-2.2$ to 1.1 mag. Such a large bias can be easily detected.
We recall that this consideration is only valid when the component is so faint that it does not affect motion of the system's photocentre.

We now estimate the bias in absolute magnitude from a substellar companion.  From Eq.~(\ref{eq:par-est-circ}) we find, using the Kepler's third law~(\ref{eq:kepler}), that
\begin{equation}
 \begin{aligned}
 \Delta M&=
 M_\rmn{est}-M=5\lg\frac{\varpi_\rmn{est}}{\varpi}=5\lg\left(1-\rho\frac{a_\star}{A}\right)\\
 &\approx -2\rho\left(\frac{M_p}{M_J}\right)\left(\frac{M_\star}{M_{\sun}}\right)^{-2/3}\,\rmn{mmag}\,,
 \end{aligned}
\end{equation}
where $M_J$ stands for the Jupiter mass. We note that $\Delta M$ is independent of distance. This relation implies that the bias $\Delta M$ is at a level the a few mmag, or less, for companions of planetary mass. Such a small luminosity difference is of limited relevance because it is well below the uncertainties in stellar calibration models. Massive brown dwarfs with $M_p\simeq$\,50--100\,$M_J$, for which the magnitude bias may amount to 0.1--0.2\,mag, offer better chance of recognizing the discrepancy in luminosity. Thus, the possibility of breaking the degeneracy between the orbital and parallactic motion using photometric data is limited to brown dwarfs, whereas detection of effects from planetary companions is hardly achievable in practice.

\section{Conclusions}
\label{s:conclusions}

We present an analysis of the effects of the Earth orbital motion on astrometric detectability of systems comprising unseen companions, with an emphasis on exoplanet detection. We demonstrated that, if period of a companion is close to one year and its orbital plane is nearly parallel to the ecliptic, orbital motion of the host may be entirely or partially absorbed by the parallax parameter. If full absorption occurs, the companion is astrometrically undetectable.

Analysis of the goodness-of-fit $\chi^2$ statistic for parallax estimation enabled us to find accompanying parallax bias and to introduce a convenient measure for detectability, effective astrometric signature, which accounts for the Earth orbital motion and effect of the parallax absorption on astrometric residuals. Remarkably, the effects of the Earth orbital motion are conveniently parametrized by one parameter: the position correlation function $\rho$. Considering circular orbit, we obtained general expression for the correlation function in terms of orbit parameters and duration of observations. Thus, we provide a complete set of formulae for calculation of the astrometric effects due to interaction between the orbital motion of the Earth and orbital motion of the host star for arbitrary circular orbits. These effects are significant for orbits with low eccentricity ($e\la0.5$) and period $0.8\loa P\loa 1.2$\,yr. In this period range, the effects crucially depend on the inclination of the orbital plane with respect to the ecliptic and on the relative position of the Earth and the star.

Some astrometrically undetectable systems can be discovered with other detection techniques. While photometric transit searches are useful for these purposes only within a narrow strip around the ecliptic, radial velocity surveys are potentially capable of detecting such systems on much of the celestial sphere, except areas around the ecliptic poles. Moreover, discrepancy between the host star luminosity derived from trigonometric parallax and astrophysical data can provide evidence for an unseen companion. Although small parallax bias due to planetary companions slightly affects absolute magnitude estimations, this effect can be significant for massive brown dwarfs companions.

It is worth mentioning that all results obtained for a single planet hold for multiple exoplanet systems because host star motion is a simple composition of effects produced by each of the planets individually. Effects of the parallax absorption on characterization of planetary orbits were not considered in this work; these complex and important problem deserves a special study.

\section*{Acknowledgements}

The authors is very much indebted to the referee, Berry Holl (Geneva Observatory), for his valuable comments and suggestions that have greatly improved an initial version of the paper. The author is grateful to the Lohrmann Observatory, Technische Universit\"at Dresden, where this work was initiated, and acknowledges support from the Deutsche Zentrum f\"ur Luft- und Raumfahrt e.V. (DLR). The author also thanks Lennart Lindegren, Fran\c{c}ois Mignard and Frederic Arenou for useful discussions. This research has made use of NASA's Astrophysics Data System.


\bibliographystyle{mnras}
\bibliography{exo}

%%%%%%%%%%%%%%%%% APPENDICES %%%%%%%%%%%%%%%%%%%%%

\appendix

\section{Goodness-of-fit for parallax estimation}
\label{app:fitting}

The procedure for fitting of the astrometric and Keplerian models to astrometric data is described in various publications  \citep[e.g.,][]{Casertano+2008,Wright+Howard2009,Perryman+2014,2017arXiv1704.02493}. The objective of this appendix is to derive $\chi^2$ statistic which serves to measure the goodness-of-fit for parallax estimation and to consider how it is affected by orbital motion of the host star.

In subsequent equations let $\bar{\bmath{u}}$ denote the direction, which takes account of all the effects except parallax. Linearization of Eqs.~(\ref{eq:uStar}) and (\ref{eq:uExo}) with respect to the parallax yields the expected direction
\begin{equation}
  \bmath{u}_\rmn{calc}=\bar{\bmath{u}}+
  \bar{\bmath{u}}\bmath{\times}\left[\bar{\bmath{u}}\bmath{\times}\varpi_\rmn{est}\bmath{b}/A\right]
\end{equation}
and the observed direction
\begin{equation}
  \bmath{u}_\rmn{obs}=\bar{\bmath{u}}+
  \bar{\bmath{u}}\bmath{\times}
  \left[
   \bar{\bmath{u}}\bmath{\times}\varpi\left(\bmath{b}-\bmath{s}\right)/A
  \right]\,.
\end{equation}
Here $\varpi_\rmn{est}$ and $\varpi$ signify the estimated and true values of the trigonometric parallax, respectively.

The least-squares solution is equivalent to the minimization of the chi-squared statistic, which measures the goodness-of-fit \citep[see, for example,][]{brandt,bevington+robinson},
\begin{equation}\label{eq:chi2-definition}
  \chi^2\left(\varpi_\rmn{est}\right)=\sum_iw_i\left(\Delta\bmath{u}_i\right)^2\,,
\end{equation}
where $i$ is the index of observation, $w$ is the statistical weight of the observation and $\Delta\bmath{u}$ is the residual in the directions,
\begin{equation}
  \Delta\bmath{u}=\bmath{u}_\rmn{obs}-\bmath{u}_\rmn{calc}\,.
\end{equation}
A straightforward calculation gives
\begin{equation}
  \left(\Delta\bmath{u}\right)^2=
  \left(
   \varpi\left(\bmath{b}-\bmath{s}\right)/A-\varpi_\rmn{est}\bmath{b}/A
  \right)^2\,
  \sin^2\nu\,,
\end{equation}
where $\nu$ denotes the angle between $\bar{\bmath{u}}$ and the vector expression $\varpi\left(\bmath{b}-\bmath{s}\right)-\varpi_\rmn{est}\bmath{b}$. The factor $\sin^2\nu$ takes account of the orbit projection on the tangent plane. This factor can be shown to slightly affect the parallax estimate and therefore is omitted in what follows.

Using the angular brackets to signify the weighted averaging,
\begin{equation}
 \left\langle f\right\rangle=
 \sum_iw_if_i\,,
\end{equation}
we can write the chi-squared statistic (\ref{eq:chi2-definition}) as
\begin{equation}\label{eq:chi2}
  \chi^2\left(\varpi_\rmn{est}\right)=
  \left\langle\left(\varpi\left(\bmath{b}-\bmath{s}\right)/A-\varpi_\rmn{est}\bmath{b}/A\right)^2\right\rangle\,.
\end{equation}
It is worth noting that, because of the omission of the factor $\sin^2\nu$, this formula gives upper limit for the goodness-of-fit. For analytical calculations, it is convenient to replace the observation-averaged quantities with time-averaged ones,
\begin{equation}\label{eq:time-average}
 \left\langle f\right\rangle=
 \frac{1}{T}\int_0^Tf\left(t\right)\rmn{d}t\,,
\end{equation}
with $T$ being the duration of observations. This change from summation to integration provides a good approximation in the practically important case of uniformly distributed observations of equal weight. It may be called the continuous observation approximation.

\section{Mutual orientation of coordinate grids (normal triads)}
\label{app:angle}

\begin{figure}
  \centering
  \includegraphics[width=\hsize]{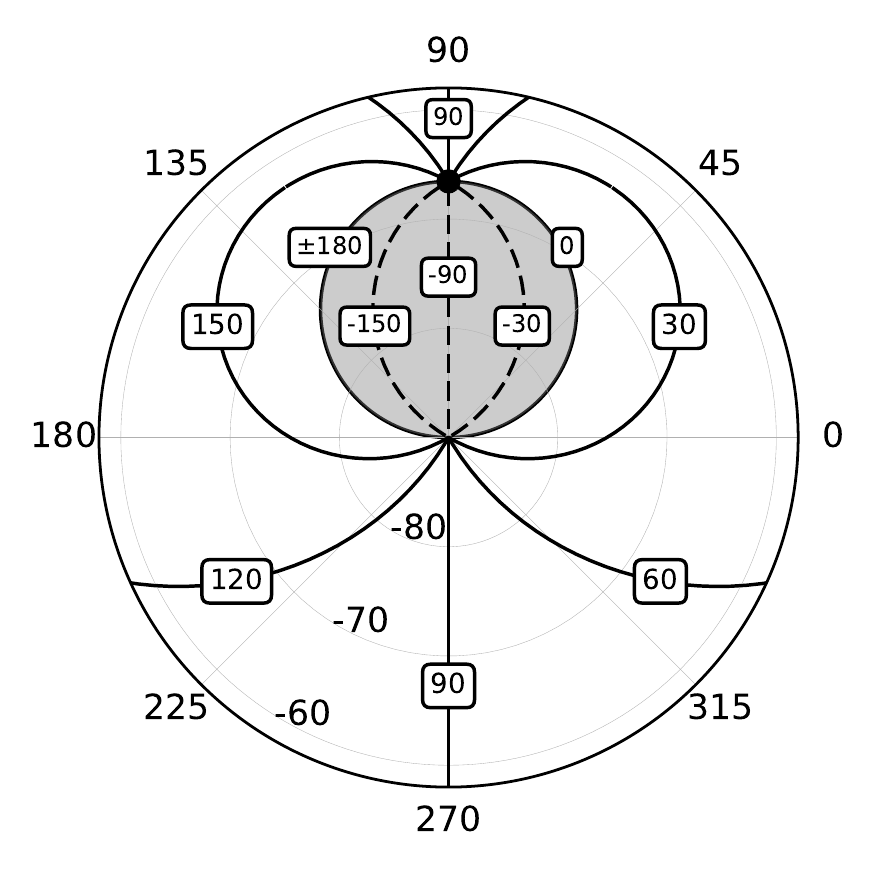}
 \caption{Distribution of the angle $\theta$ calculated from Eqs.~(\ref{eq:app-cos-theta}) and (\ref{eq:app-sin-theta}) for $\delta<{-60}\degr$. The bold lines delineate constant value of $\theta$ indicated by corresponding labels. The centre of the plot is situated at the south celestial poles ($\delta=-90\degr$), while the black circle corresponds to the south ecliptic pole ($\beta=-90\degr$). The numbers next to the coordinate grid lines give right ascension and declination. Negative values of  $\theta$ are contained within the shaded area.}
  \label{fig:polar}
\end{figure}

In the equatorial coordinates, local directions of increasing $\alpha$ and $\delta$ are specified by the unit vectors $\bmath{p}$ and $\bmath{q}$.
Together with the direction vector $\bmath{r}$, they constitute the right-handed orthogonal vector triad, commonly referred to as the normal triad relative to the equatorial system \citep{lindegren2013}. Its coordinate representation is
\begin{equation}
  \begin{bmatrix}
    \bmath{p} & \bmath{q} & \bmath{r}
  \end{bmatrix}
  =
  \begin{bmatrix}
    -\sin\alpha & -\cos\alpha\sin\delta & \cos\alpha\cos\delta \\
     \cos\alpha & -\sin\alpha\sin\delta & \sin\alpha\cos\delta \\
     0          &            \cos\delta & \sin\delta
  \end{bmatrix}\,.
\end{equation}
Similarly, the unit vectors $\bmath{p}_\rmn{E}$ and $\bmath{q}_\rmn{E}$ specify the directions of increasing $\lambda$ and $\beta$ in the ecliptic coordinates. The coordinate representation of the corresponding normal triad is
\begin{equation}
  \begin{bmatrix}
    \bmath{p}_\rmn{E}^\rmn{ec} & \bmath{q}_\rmn{E}^\rmn{ec} & \bmath{r}^\rmn{ec}
  \end{bmatrix}
  =
  \begin{bmatrix}
    -\sin\lambda & -\cos\lambda\sin\beta & \cos\lambda\cos\beta \\
     \cos\lambda & -\sin\lambda\sin\beta & \sin\lambda\cos\beta \\
     0           &             \cos\beta &            \sin\beta
  \end{bmatrix}\,.
\end{equation}
Here the superscript `ec' indicates that the components of the vectors refer to the axes of the ecliptic system. The transformation to the equatorial system is done by means of a rotation matrix:
\begin{equation}
  \begin{bmatrix}
    \bmath{p}_\rmn{E} & \bmath{q}_\rmn{E} & \bmath{r}
  \end{bmatrix}
  =
  \begin{bmatrix}
    1 & 0            & 0             \\
    0 & \cos\epsilon & -\sin\epsilon \\
    0 & \sin\epsilon &  \cos\epsilon
  \end{bmatrix}
    \begin{bmatrix}
    \bmath{p}_\rmn{E}^\rmn{ec} & \bmath{q}_\rmn{E}^\rmn{ec} & \bmath{r}^\rmn{ec}
  \end{bmatrix}\,,
\end{equation}
with $\epsilon$ being the obliquity of the ecliptic.

In the tangent plane, the vector pairs $\bmath{p}$, $\bmath{q}$ and $\bmath{p}_\rmn{E}$, $\bmath{q}_\rmn{E}$ are rotated relative to each other through some angle. Since the vectors are along to relevant coordinate lines, this angle is a measure of mutual orientation of equatorial and ecliptic coordinate grids at a given point on the celestial sphere. As explained in Sect.~\ref{ss:orientation}, for our purposes it is convenient to specify this rotation by the angle $\theta$ counted counterclockwise from $\bmath{q}$ to $\bmath{p}_\rmn{E}$. The relation between the vector pairs can then be written as
\begin{align}\label{eq:pq-pairs}
  \begin{bmatrix}
    \bmath{p} & \bmath{q}
  \end{bmatrix}
  &=
  \begin{bmatrix}
    \bmath{p}_\rmn{E} & \bmath{q}_\rmn{E}
  \end{bmatrix}
  \begin{bmatrix}
     \sin\theta & \cos\theta \\
    -\cos\theta & \sin\theta \\
  \end{bmatrix}\,.
\end{align}
Since value of $\theta$ lies between $-\upi$ and $\upi$, it is uniquely specified by its sine and cosine, which are computed from the relevant dot products:
\begin{align}
  \label{eq:app-cos-theta}
  \cos\theta &=\bmath{q}\bmath{\cdot}\bmath{p}_\rmn{E}=-\bmath{p}\bmath{\cdot}\bmath{q}_\rmn{E}\,,\\
  \label{eq:app-sin-theta}
  \sin\theta &=\bmath{p}\bmath{\cdot}\bmath{p}_\rmn{E}=\bmath{q}\bmath{\cdot}\bmath{q}_\rmn{E}\,.
\end{align}
These formulae determine the angle $\theta$ at any position. Omitting intermediate calculations, we give the final results in terms of the equatorial coordinates:
\begin{align}
 \sin\theta&=
  \frac{\cos\epsilon\cos\delta+\sin\epsilon\sin\delta\sin\alpha}
       {\sqrt{1-\left(\cos\epsilon\sin\delta-\sin\epsilon\cos\delta\sin\alpha\right)^2}}\,,\\
 \cos\theta&=
  \frac{\sin\epsilon\cos\alpha}
       {\sqrt{1-\left(\cos\epsilon\sin\delta-\sin\epsilon\cos\delta\sin\alpha\right)^2}}\,.
\end{align}
On much of the celestial sphere, $\theta$ is positive. For example, it varies around the celestial equator ($\delta=0$) between limits of $\upi/2-\epsilon$ and $\upi/2+\epsilon$. Near the poles, where $\delta$ or $\beta$ are close to $\pm\upi/2$, it can take any value in the range from $-\upi$ to $\upi$. Figure~\ref{fig:polar} illustrates how $\theta$ behaves in the region close to the south poles.

\section{Evaluation of means for an elliptic orbit}
\label{app:ellipse}

This appendix gives derivations of the means $\left\langle s^2\right\rangle$ and $\left\langle\bmath{s\cdot b}\vphantom{s^2}\right\rangle$ for a host star on an elliptic orbit on the assumptions described in Sect.~\ref{ss:ellipse}.
Barycentric distance of the host star is
\begin{equation}
  s=a_\star\left(1-e\cos E\left(t\right)\right)\,.
\end{equation}
The relation between the eccentric anomaly $E$ and time is given by Kepler's equation. For $P=1\,\rmn{yr}$ and pericentre passage occurring at $t=0$, this relation takes the form
\begin{equation}
  2\upi t=E\left(t\right)-e\sin E\left(t\right)\,.
\end{equation}
The mean value of $s^2$ over the orbital period is
\begin{equation}
  \left\langle s^2\right\rangle=a_\star^2\int_0^1\left[1-e\cos E\left(t\right)\right]^2\rmn{d}t\,.
\end{equation}
Transforming from integral over $t$ to one over $E$, we have
\begin{equation}\label{eq:app-ell-s2}
  \left\langle s^2\right\rangle
  =\frac{a_\star^2}{2\upi}\int_0^{2\upi}\left(1-e\cos E\right)^3\rmn{d}E
  =a_\star^2\left(1+\frac{3}{2}e^2\right)\,.
\end{equation}
For a zero initial orbital phase, the position of Earth is given by Eq.~(\ref{eq:b}) with $\psi=0$:
\begin{equation}
 \bmath{b}=
 \begin{pmatrix}
    A\cos 2\upi t, &
    A\sin 2\upi t, &
    0
 \end{pmatrix}\,.
\end{equation}
It is known from the theory of elliptic motion that position of the host star can be written in terms of the eccentric anomaly as
\begin{equation}
 \bmath{s}=
 \begin{pmatrix}
    a_\star\left(\cos E-e\right), &
    a_\star\sqrt{1-e^2}\sin E, &
    0
 \end{pmatrix}\,.
\end{equation}
This equation takes into account the assumption that the orbital plane is parallel to the ecliptic, $i_\rmn{E}=0$.

The mean of the dot product $\bmath{s\cdot b}$ over the orbital period is
\begin{equation}\label{eq:app-ell-cov}
 \begin{aligned}
  \left\langle\bmath{s\cdot b}\right\rangle&=a_\star A\,\times\\&\int_0^1\left[\left(\cos E-e\right)\cos 2\upi t+\sqrt{1-e^2}\sin E\sin 2\upi t\right]\rmn{d}t\,.
 \end{aligned}
\end{equation}
If we again transform to integral over $E$, the integrand becomes
\begin{equation}
 \begin{aligned}
  &\left[\left(\cos E-e\right)\cos\left(E-e\sin E\right)\vphantom{\sqrt{e^2}}\right.\\
  &+\left.\sqrt{1-e^2}\sin E\sin\left(E-e\sin E\right)\right]\times\left(1-e\cos E\right)\,.
 \end{aligned}
\end{equation}
Simple, though lengthy, calculations show that this expression can be written as
\begin{equation}
 \begin{aligned}
  &-\frac{3}{2}e\cos\left(E-e\sin E\right)\\
  &+\frac{1+e^2}{2}\left[\cos\left(e\sin E\right)+\cos\left(2E-e\sin E\right)\right]\\
  &-\frac{e}{4}\left[\cos\left(E+e\sin E\right)+\cos\left(3E-e\sin E\right)\right]\\
  &+\left\{\frac{1}{2}\left[\cos\left(e\sin E\right)-
                            \cos\left(2E-e\sin E\right)
                     \right]\right.\\
  &\left.-\frac{e}{4}\left[\cos\left(E+e\sin E\right)-
                           \cos\left(3E-e\sin E\right)
                     \right]\right\}
  \sqrt{1-e^2}\,.
 \end{aligned}
\end{equation}
Integrals of such terms over $E$ from 0 to $2\upi$ are calculated by using a formula of the theory of Bessel functions,
\begin{equation}
  \frac{1}{2\upi}\int_0^{2\upi}\cos\left(nE-e\sin E\right)\rmn{d}E=J_n\left(e\right)\,,
\end{equation}
where $J_n\left(e\right)$ is the Bessel function of order $n$. Applying this formula, we find that the integral in Eq.~(\ref{eq:app-ell-cov}) equals to
\begin{equation}
 \begin{aligned}
  &-\frac{3}{2}eJ_1\left(e\right)\\
  &+\frac{1+e^2}{2}\left[J_0\left(e\right)+J_2\left(e\right)\right]
  -\frac{e}{4}\left[J_1\left(-e\right)+J_3\left(e\right)\right]\\
  &+\left\{\frac{1}{2}\left[J_0\left(e\right)-J_2\left(e\right)
                     \right]
  -\frac{e}{4}\left[J_1\left(-e\right)-J_3\left(e\right)
                     \right]\right\}
  \sqrt{1-e^2}\,.
 \end{aligned}
\end{equation}
Using the well-known properties of the Bessel functions,
\begin{align}
  J_1\left(-e\right) &= -J_1\left(e\right)\,, \\
  J_2\left(e\right) &= \frac{2}{e}J_1\left(e\right)-J_0\left(e\right)\,, \\
  J_3\left(e\right) &= \left(\frac{8}{e^2}-1\right)J_1\left(e\right)-\frac{4}{e}J_0\left(e\right)\,,
\end{align}
we finally obtain the following expression for the required mean value
\begin{equation}\label{eq:app-ell-sb}
  \left\langle\bmath{s\cdot b}\right\rangle=a_\star A\left(J_0\left(e\right)-\frac{1-\sqrt{1-e^2}}{e}J_1\left(e\right)\right)\,.
\end{equation}
For small eccentricity, this formula is approximated by
\begin{equation}\label{eq:app-ell-sb-approx}
  \left\langle\bmath{s\cdot b}\right\rangle=a_\star A\left(1-\frac{e^2}{2}\right)\,.
\end{equation}
Thus, both the means, $\left\langle s^2\right\rangle$ and $\left\langle\bmath{s\cdot b}\vphantom{s^2}\right\rangle$, quadratically depend on eccentricity when orbit is nearly circular. In an extreme case of very elongated orbit ($e\to 1$) they tend to the following limiting values:
\begin{align}
  \left\langle s^2\right\rangle &\to \frac{5}{2}a_\star^2\,, \\
  \left\langle\bmath{s\cdot b}\vphantom{s^2}\right\rangle &\to a_\star A\left(J_0\left(1\right)-J_1\left(1\right)\right)\approx 0.33\,a_\star A\,.
\end{align}
%%%%%%%%%%%%%%%%%%%%%%%%%%%%%%%%%%%%%%%%%%%%%%%%%%

% Don't change these lines
\bsp	% typesetting comment
\label{lastpage}
\end{document}